# An Empirical Study on Content Bundling in BitTorrent Swarming System

Jinyoung Han
Seoul National University
jyhan@mmlab.snu.ac.kr

Taejoong Chung
Seoul National University
tjchung@mmlab.snu.ac.kr

Seungbae Kim
Seoul National University
sbkim@mmlab.snu.ac.kr

Hyun-chul Kim
Seoul National University
hkim@mmlab.snu.ac.kr

Ted “Taekyoung” Kwon
Seoul National University
tkkwon@snu.ac.kr

Yanghee Choi
Seoul National University
yhchoi@snu.ac.kr

## ABSTRACT

Despite the tremendous success of BitTorrent, its swarming system suffers from a fundamental limitation: lower or no availability of unpopular contents. Recently, Menasche *et al.* has shown that *bundling* [1] is a promising solution to mitigate this availability problem; it improves the availability and reduces download times for unpopular contents by combining multiple files into a single swarm. There also have been studies on bundling strategies and performance issues in bundled swarms. In spite of the recent surge of interest in the benefits of and strategies for bundling, there are still little empirical grounding for understanding, describing, and modeling it. This is the first empirical study that measures and analyzes how prevalent contents bundling is in BitTorrent and how peers access the bundled contents, in comparison to the other non-bundled (i.e., single-filed) ones. To our surprise, we found that around 70% of BitTorrent swarms contain multiple files, which indicate that bundling has become widespread for contents sharing. We also show that the amount of bytes shared in bundled swarms is estimated to be around 85% out of all the BitTorrent contents logged in our datasets. Inspired from our findings, we raise and discuss three important research questions in the field of file sharing systems as well as future contents-oriented networking: i) bundling strategies, ii) bundling-aware sharing systems in BitTorrent, and iii) implications on content-oriented networking.

## Keywords

Content Bundling, Swarming System, Peer-to-Peer System, BitTorrent, Content-Oriented Networking

## 1. INTRODUCTION

BitTorrent [7] has become the de facto standard in sharing contents on the Internet. According to the Ipoque's report released in 2009 [2], BitTorrent accounts for approximately 27-55% of today's Internet traffic. The ever increasing usage of BitTorrent is attributed to some attractive properties of its swarming systems: First, cooperation among peers in a swarm stimulated by the tit-for-tat based incentive mechanism to improve the overall system performance in terms of throughput. Second, the tit-for-tat strategy also prevents (or mitigates) the free-riding problem. Third, the swarming technique scales well even in the presence of massive flash crowds for popular contents [15, 18].

Despite the success of BitTorrent, its swarming system suffers from a fundamental limitation: little or no availability of unpopular contents. That is, peers arriving after the initial flash crowd may often end up with finding the content unavailable, not to mention unpopular contents [17]. Recently, Menasche *et al.* has shown that *bundling* is a promising solution to mitigate this availability problem; it improves the availability and reduces download times for unpopular contents by combining multiple files into a single swarm [17]. There also have been studies on bundling strategies [11, 16] and performance issues in bundled swarms [21].

In spite of the recent surge of interest in the benefits of and strategies for bundling, there is still little empirical grounding for understanding, describing, and modeling it. This leaves researchers uncertain of (i) whether and how the assumptions (on both bundling pattern and user access pattern [17, 21], with little if any empirical evidence) made in the aforementioned work are congruent with (or close to) reality, and (ii) the status quo and performance and architectural implications of bundling for the advancement of file sharing techniques or architectures (not to mention BitTorrent), as well as of the so-called future *‘content-oriented networking’*.

To begin to provide empirical grounds for understanding, describing, and modeling contents bundling in BitTorrent, we make the following contributions: (1) This is the first empirical study that measures and analyzes how prevalent contents bundling is in BitTorrent and how peers access the bundled contents, in comparison to the other non-bundled (i.e., single-filed) ones. To this end, we first developed a BitTorrent monitoring software by modifying the Azureus [2] [3], one the most popular BitTorrent clients. Using the developed code, we have collected over a month of data comprised of 36K distinct swarms of 2.8 Million distinct peers. (2) To our surprise, we found that around 70% of BitTorrent swarms contain multiple files, which indicate that bundling has become widely used for contents sharing. (3) We also estimated that the amount of bytes shared in bundled swarms are larger than that in single-filed swarms: The amount of data contained in bundled swarms are about 2 times larger than single-filed ones, the number of bundled swarms are about 2.3 times in average larger than that of single-filed ones, and the average popularity of bundled swarms are 1.2 times larger than single-filed ones, which we approximately estimate that amount of bytes shared in bundled contents account for around 85% out of all the BitTorrent contents logged in our datasets. (4) According to our measurements, the bundling pattern and user access pattern are different depending on the seven BitTorrent content type categories (i.e., Movie, Porn, TV, Music, Applications, E-book, and Game). (5) Based on our findings and lessons learned, we raise and discuss three new important research questions in the area of file sharing systems as well as the future contents-oriented networking: i) bundling strategy, ii) bundling-aware swarming systems in BitTorrent, and iii) implications on content-oriented networking.

[1] Bundling is a common strategy adopted by BitTorrent publishers by which a publisher packages a number of related files (e.g., episodes of a sitcom) and disseminates them via a single larger swarm [17], instead of disseminating individual files via separate swarms.

This paper is organized as follows. After reviewing related work in Section 2, we present measurement methodology, data, and results in Section 3. Section 4 discusses on implications of content bundling and our ongoing work. Finally, we conclude this paper in Section 5.

## 2. RELATED WORK

**BitTorrent:** BitTorrent [7] has been the de facto standard for content sharing. Due to its success in the real world, many studies have been conducted to investigate BitTorrent's behavior in terms of throughput, fairness and incentive issues, revealing valuable insights into the performance aspect of the BitTorrent [15, 18–20].

**Sharing Multiple Files:** Most studies on BitTorrent had focused only on sharing and transmission of single torrents until Guo *et al.* found that 85% of users concurrently access multiple torrents [10]. Yang *et al.* [22] proposed a novel incentive mechanism for remaining as seeds in a subset of downloaded torrents when a user downloads multiple torrents. This mechanism calculates the aggregated downloading rate in a *cross-torrent* fashion in the peer selection phase, so that a node can get additional credits from providing pieces in another torrent which it participates in as a seed. These studies assume that only a single file is shared in a single swarm. In contrast, we focus on "bundling", which allows peers to download multiple files from a single swarm.

**Bundling:** Recently, Menasche *et al.* [16, 17] has shown that bundling can mitigate the availability problem by combining multiple files into a single swarm. Also Tien *et al.* studied the performance issues in sharing multiple files not only using multiple torrents but also using a bundle in BitTorrent [21]. Despite the recent surge of interest in the benefits of and strategies for bundling, there are still little empirical grounding for understanding, describing, and modeling it.

## 3. MEASUREMENT

In this section, we present an empirical analysis of BitTorrent content bundling to quantify: (1) whether and to what extent content bundling is prevalent in BitTorrent, and (2) the bundling pattern and user access pattern depending on the seven BitTorrent content type categories (i.e., Movie, Porn, TV, Music, Applications, E-book, and Game).

### 3.1 Methodology

We conducted a measurement study on real BitTorrent swarms starting from April 22, 2010 to May 29, 2010. We first developed a BitTorrent monitoring client to keep track of each swarm by modifying the Azureus [3], one of the most popular BitTorrent client. We then developed a torrent crawling agent to fetch the torrent file (i.e., a ".torrent" file) and its information from a torrent hosting site. The crawling agent periodically fetches the recently released torrent files and their information from The Pirate Bay (TPB) [5] which is the most popular torrent hosting site, and then sends the torrent information to a subset of the monitoring clients. When a monitoring client receives a torrent information from the crawling agent, it joins the swarm and begins to monitor each peer. The monitoring client requests the peer list to the tracker several times through the tracker protocol [1] so that it can get most of the peer information in a swarm. In addition, it also leverages the Peer Exchange Extension (PEX) [4] protocol to discover the rest of the peers to cover the entire swarm. We finally save all the swarm information from the torrent hosting site (i.e, TPB), the tracker, and each connected peer in the swarm. Each entry in the log data consists of torrent meta information (e.g., a torrent name, file names and a torrent creation time), information of peers (e.g., an IP address and its bitmap [3] for content pieces), and information from the torrent hosting site (e.g., categories of

[2]Azureus is also called Vuze, one of the most popular BitTorrent client.

[3]The bitmap is used to convey the blocks that a user pos-

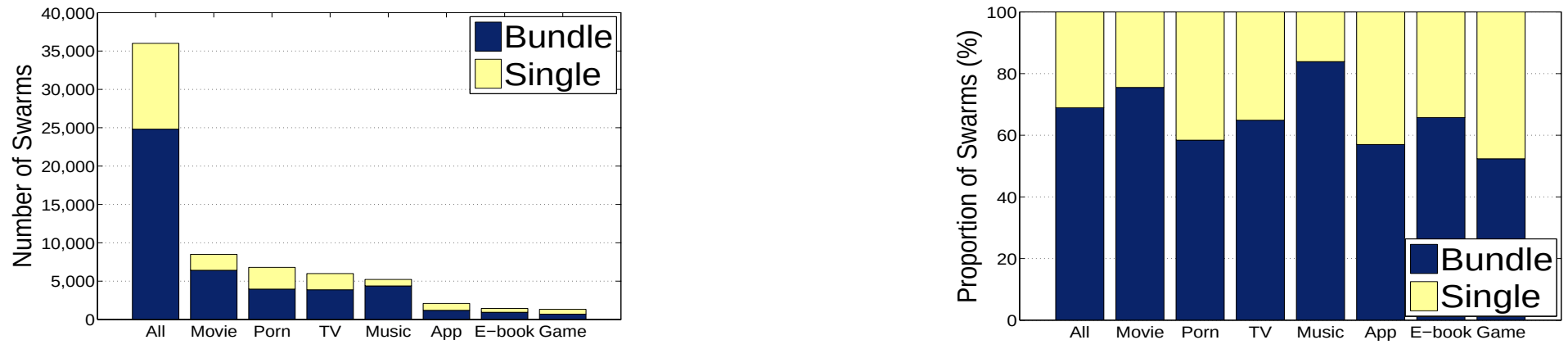

(a) Number of Swarms for Bundled and Single-filed

(b) Proportion of Swarms for Bundled and Single-filed

Figure 1: Bundling is widely used in BitTorrent.

the torrent file and release time of the torrent file). We have collected data comprised of 36K distinct swarms and 2.8 Million peers.

Throughout this paper, we investigate the bundling pattern and user access pattern based on the seven BitTorrent content type categories [4] (i.e., Movie, Porn, TV, Music, Application, E-book, Game) which accounts for 87% of the swarms logged in our datasets. The number of swarms of Movie, Porn, TV, Music, Application, E-book and Game accounts for 24%, 19%, 17%, 15%, 6%, 4%, and 4% of all the swarms, respectively.

## 3.2 Bundling is Widespread

To analyze how prevalent contents bundling is in BitTorrent and how peers access the bundled contents, we compare bundled swarms to the other non-bundled (i.e., single-filed) ones in terms of the number of swarms, swarm size (i.e., summation of file sizes in a swarm), popularity, and availability.

Figure 1 shows that around 70% of BitTorrent swarms contain multiple files, which means content bundling is widely used in BitTorrent today for content sharing. For the music type, over 80% of swarms are bundled ones, which indicates that users often share music files through bundling based on the players, composers, or albums. Meanwhile, around 50% of swarms are bundled for the types of game and application, which are lower than other content types, because games and applications are sometimes shared by a single installation file (e.g., a single ISO file or a ZIP file) and sometimes shared by multiple files which consist of multiple installation files and subsidiary files such as How-To documents.

We next compare the swarm size of bundled and single-filed swarms in Figure 2. As there are a number of files in bundled swarms while there is only a file in single-filed swarms, the swarm size of bundled ones (1.2GB in average) is typically larger than that of single-filed ones (0.6GB in average). That is, the amount of sesses to its neighbors, and it is part of the BitTorrent protocol.

[4]We used the content type categories information which are tagged in The Pirate Bay [5].

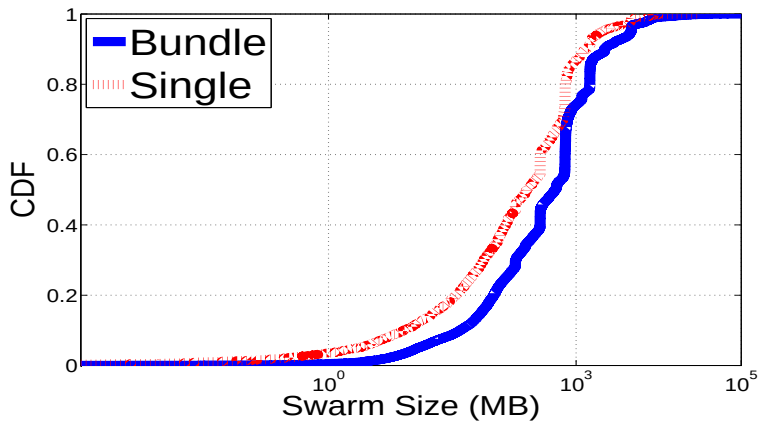

Figure 2: Amount of data contained in bundled swarms are larger than the single-filed ones.

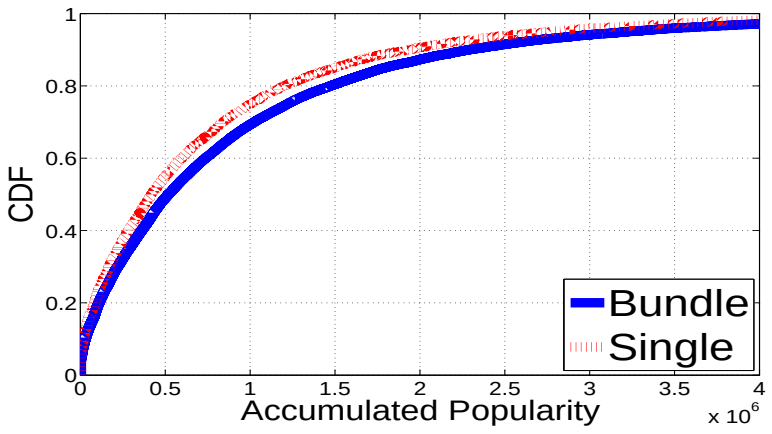

Figure 3: Bundled files are more popular than single-filed ones.

data contained in bundled swarms are typically larger than than that in single-filed ones.

To investigate and compare the usage patterns of bundled swarms and single-filed ones in BitTorrent, we measure the popularity of bundled swarms and the single-filed ones. As Dan *et al.* showed in [8], there are two methods to quantify the content popularity; instantaneous popularity and the popularity over time [5]. To measure not only the instantaneous popularity but also the popularity over time, we combine the concepts of two methods into a single novel content popularity metric, which we define "accumulated popularity". The accumulated popularity $P_A$ is calculated as follows:

$$P_A = \int_0^T N(t)dt$$

where T is the last time that we want to estimate the popularity and N(t) is the number of peer at time t. As shown in Figure 3, bundled swarms are more popular than single-filed ones (about 1.2 times larger in average).

[5]In [8], the popularity over time is defined as the number of times the contents are downloaded in a time interval, and the instantaneous popularity is defined as the number of peers that simultaneously participate in the swarm.

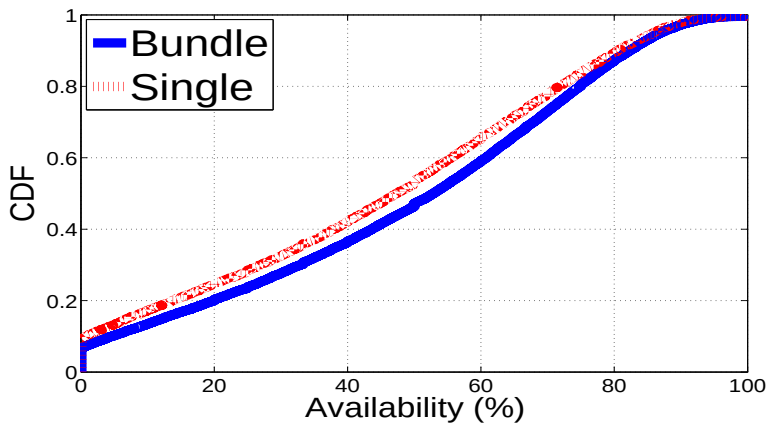

Figure 4: Bundled swarms are more available than single-filed ones.

We then analyze the availability of bundled swarms and single-filed ones in BiTorrent. While some studies such as [17] measure the just seed availability in their measurements, we measure the availability based on the content piece information of peers because the availability of a swarm in practice depends not only on the presence of seeds [6] but also on the union of pieces that leechers have in a swarm. To calculate the availability based on the content piece, we combine all the piece maps of peers in a swarm, and then we calculate the portions of pieces which are available in the swarm. Figure 4 shows that bundled swarms are more available than single-filed ones (the average availability in bundled swarms is 49% while 44% in single-filed ones).

Based on above measurements and analysis, we estimate that the amount of bytes shared in bundled swarms are much larger than that in single-filed ones: The amount of data contained in bundled swarms are about 2 times larger than single-filed ones, the number of bundled swarms are about 2.3 times larger than that of single-filed ones, and the average popularity of bundled swarms are 1.2 times larger than single-filed ones, which we approximately estimate that amount of bytes shared in bundled contents account for around 85% out of all the BitTorrent contents logged in our datasets.

### 3.3 Bundling Pattern and User Access Pattern Per Content Type

In this subsection, we investigate the bundling pattern and user access pattern depending on the seven content type categories.

Figure 5 shows the comparisons between bundled swarms and single-filed ones in terms of the popularity, availability, and swarm size depending on the seven content type categories. Bundled swarms are more popular and available than single-filed ones in most content types in Figures 5a and 5b. More specifically, bundled swarms of the Movie, Porn, Music, and Game are more popular and available than single-filed ones, while the popularity and availability of bundled swarms of the E-book, Application, and TV are almost similar or sometimes lower than single-filed ones. Note that bundled swarms of the Game are noticeably popular, which indicates that though the number of bundled swarms made by publishers is almost similar with single-filed ones as in 1, users prefer to access or download the bundled swarms more.

Figure 5c shows that the swarm size of the bundled swarms are much larger than that of single-filed ones in all the content types. Especially, the swarm size of bundled swarms of the Game are around 2.8GB in average which is more than 3 times bigger than other types.

Figure 6 shows the number of files in a bundle and the proportion of the number of files selected and downloaded by users from a bundle depending on the content categories. We first show the CDF of the number of files in a bundle in Figure 6a. As shown in Figure 6a, bundled swarms of the Music contain more files than other content types (80% of bundled swarms of the Music contain more than 10 files). On the other hand, for the E-book, 90% of bundled swarms contain less than 4 files. Around 85% of bundled swarms of the Movie and the Application, 75% of bundled swarms of the TV, and 70% of bundled swarms of the Game contain less than 10 files. Note that around 70% of bundled swarms of the Porn contain less than 7 files while over 15% of bundled swarms of the Porn contain over 100 files.

Figure 6b shows the CDF of ratio of files selected (or requested) by users in a bundled swarm. In current BitTorrent, there is a procedure that users can select a set of files among bundled files in a swarm. While, though users have the options to choose the files, over 90% files in a bundled swarm are selected by users in Figure 6b, which indicates that the bundled contents composed by uploaders (i.e., publishers) are usually accepted (or agreed) by downloaders.

We then examine the correlation of files in a bundled swarm. If a user request two fils simultaneously, then we assume that the requested two files are correlated. Based on this assumption, we calculate the correlation of two files by measuring user access history. Figure 6c shows the CDF of the correlation of files in a bundle. Because users usually make bundled swarms by combining multiple related files, files in a bundle are highly correlated (over 90% in average) as shown in Figure 6c.

## 4. DISCUSSION

Inspired from our measurements, we raise and discuss three important research questions in the field of file sharing techniques (or architectures) as well as future contents-oriented networking:

- Our measurement study shows that content bundling

[6]Peers that possess all pieces in the content are called seeds while peers that have not ye completed their downloads are called leechers.

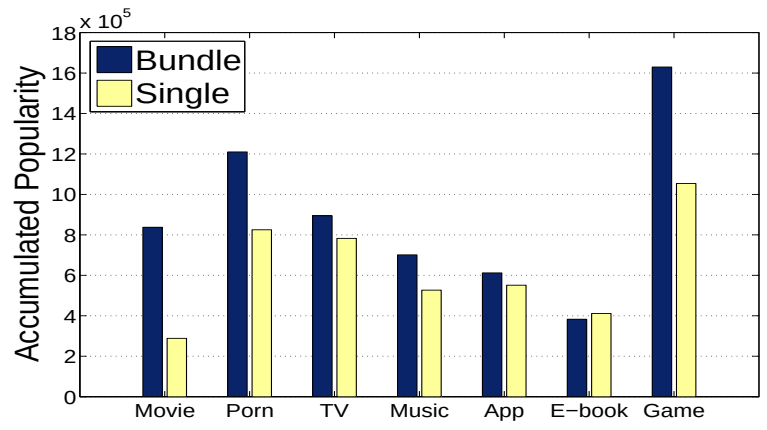

(a) Average Accumulated Popularity

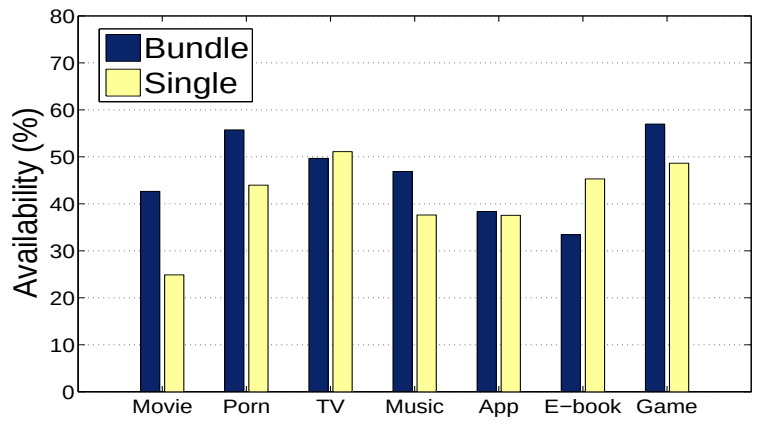

(b) Average Availability

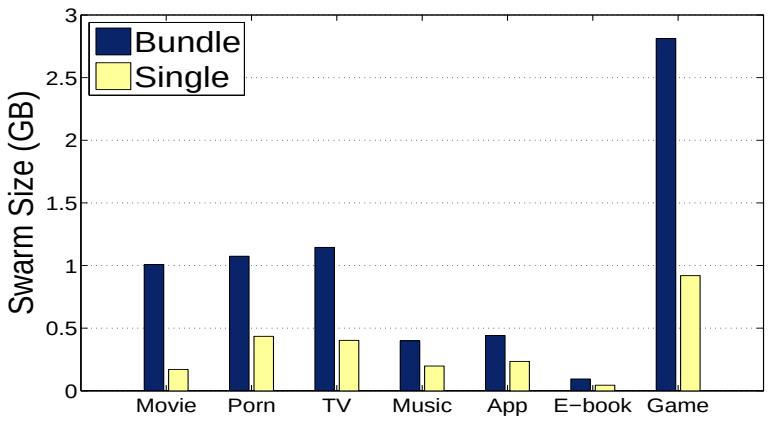

(c) Average Swarm Size

Figure 5: Comparisons between bundled swarms and single-filed ones depending on the content category

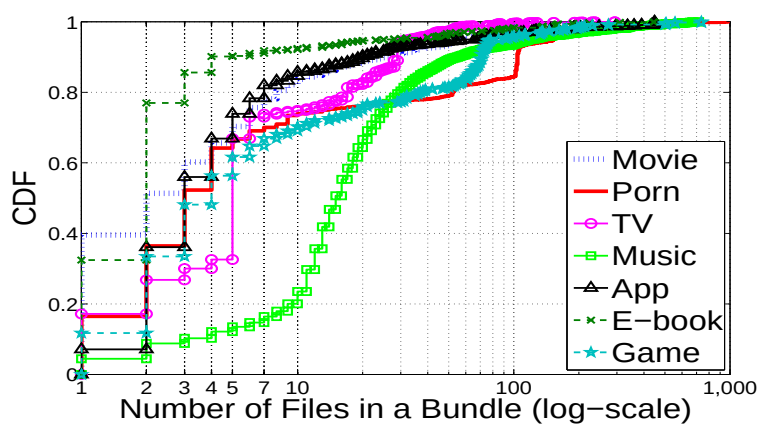

(a) Number of File in a Bundled Swarm

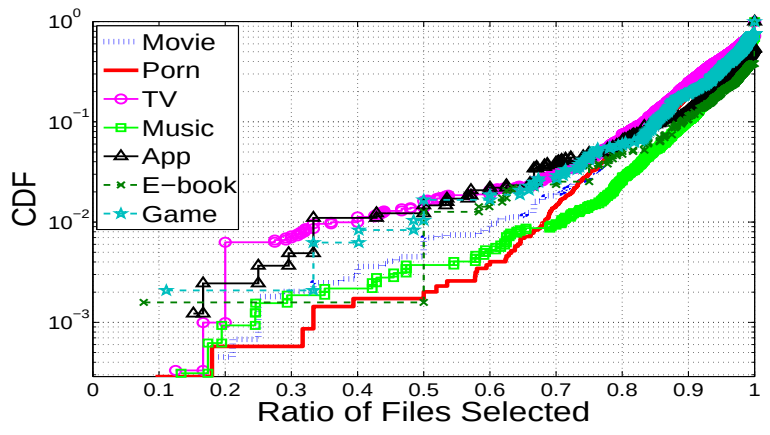

(b) Ratio of Number of Files Selected

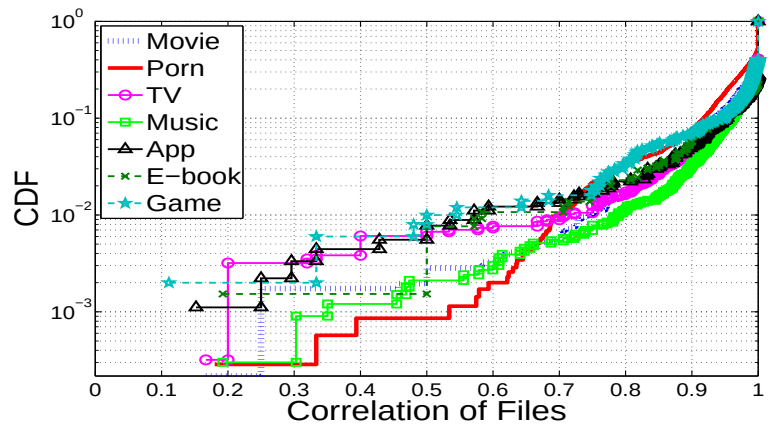

(c) Correlation of Files in a Bundled Swarm

Figure 6: Status quo of bundled contents depending on seven categories

has become widely used for contents sharing, which reveals the tendency of users to share a set of related files as a whole (i.e., bundled contents), not just as single individual files (i.e, single-filed contents). In this sense, what lessons can we learn from our findings? What are the implications of the findings on content-oriented networking?

- So far, content bundling in BitTorrent has been done manually by publishers in an ad hoc manner. If bundling is supported systematically in an automatic and efficient fashion, the system may enhance the availability and download speed. What information can be exploited to make a bundle? What should be considered to make a bundle efficiently?
- Is the current sharing mechanism for bundled contents in BitTorrent efficient? How can we maximize the performance of the bundled contents by taking the correlation of contents and cooperations of users into consideration?

**Implications on content-oriented networking:** It is widely recognized that a single network address (i.e., IP) with an identity and a location information may not meet today's content (or data) intensive application needs (e.g., High-Definition (HD) video files). There recently have been several proposals that aim to switch from host-oriented paradigm to content-oriented networking [6, 9, 12, 14]. Most studies have been conducted in naming, resolution and dissemination issues in sharing a single file on the content-oriented network, which assume that a content is regarded as a single file. That is, these content sharing systems presume the *file-oriented* paradigm, which means only a file is considered to search and share efficiently. However, it is questionable that the content in content-oriented networking should be just considered as a single file, because our measurement results show that content bundling in BitTorrent has become widely used for content sharing, which means users may prefer to share not only a file but also a set of related files (i.e,. bundle). For example, users often share songs of a recently released album, or series of famous TV shows such as "The Oprah Winfrey Show". In this case, a bundle of recently released album or a bundle of the TV show can be a content. Therefore, we believe that a bundle (or a set of files) also can be considered as a content in content-oriented networking. In this sense, we raise the problem of *file-oriented* paradigm in content-oriented networking.

Our simple study using 80 bundled contents in our measurement data shows the limitation or problem of the *file-oriented* paradigm in content-oriented networking. In our study, we calculate the similarities among files in a bundle in two ways (i.e., *(file) name-based similarity* and *attribute-based similarity*) to compare with the correlation of files based on the access history which is described in Section 3. Among multiple candidate criteria to estimate the file similarity such as file size, category (e.g., movie, TV show or game), and file hash, we first choose the file name to calculate the similarity be-cause the file name is exactly what users use to search, download, and upload files. To calculate the similarity of file names, we adopt a popular text classification algorithm: Levenshtein distance [7]. We also calculate the attribute-based similarity by making 28 attributes (e.g., a director, a distributing agency, a main actor, and etc.) for each file using a search engine (i.e., IMDB), and then we calculate the similarity among files based on the attribute information. Figure 7 shows the CDF of two similarities and the correlation of our sample bundled contents. The correlations of 60% of bundled contents are close to 1, which means files in those bundles are closely correlated. For example, a single big file can be divided into multiple files, or a video file, a sample file, and a information file of a same movie can be combined into a single content. The attribute-based similarity classifies the correlated content as well as the correlation. However, the name-based similarity cannot classify the correlation of files well because file-oriented information such as the name is limited to classify. For the system that need classify (or identify) the correlated content such as a caching system in content-oriented networking, the *file-oriented* view may not be effective. We believe that our findings raise the important research issue on the problem of *file-oriented* view in content-oriented networking.

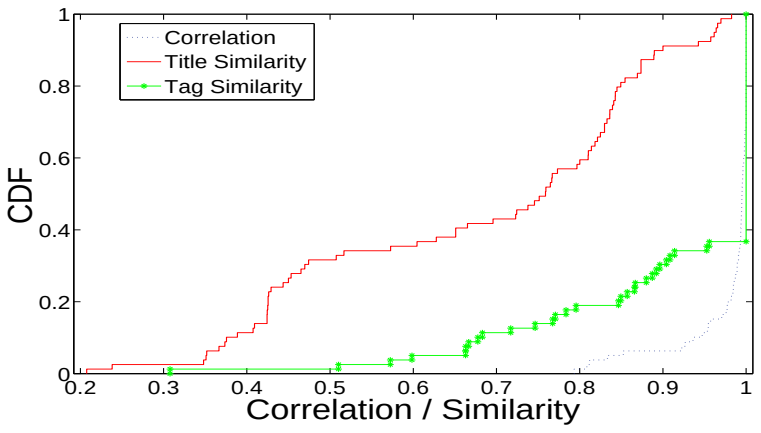

Figure 7: Sometimes, a single file has some limitations to identify/classify its content.

**How to make a bundle?:** To answer the second question, we are studying bundling strategies to combine multiple files into a bundle considering correlations of files. In our strategies, we exploit two correlations: (1) in-content attributes such as a title, topic, tag, and context of the content, (2) out-of-content attributes such as user access pattern and history to the content. Based on the above correlation information, a bundle can be made. Bundling unpopular contents as discussed in [17] is one example of using an out-of-content attribute.

**How to share a bundle?:** To address the final question, we are studying a sharing mechanism to maximize the performance of the bundle. When a bundle is shared in BitTorrent, a user can decide whether they receive all the files or some of them which are chosen by users. To accelerate cooperations among users in sharing a bundle, a user can cache and share unwanted files or already downloaded files. In our mechanism, we use a *cross-file* fashion to give incentive for the cooperations and contributions of users because files in a bundle are mostly highly correlated. This incentive mechanism is similar to the *cross-torrent* manner in [22].

[7] Given space limit, readers refer to [13] for details.

## 5. CONCLUSION

Content availability is a serious problem in today's peer-to-peer swarming systems. Menasche *et al.* has shown that bundling is a promising solution to mitigate this availability problem, but there are still little empirical grounding for understanding, describing and modeling it. This is the first empirical study that measure and analyze how prevalent contents bundling is in BitTorrent and how peers access the bundled contents, in comparison to the single-filed ones. To our surprise, we found that bundling has become widespread for contents sharing. Inspired from our findings, we raise and discuss three important research questions in the field of file sharing techniques/architecture as well as future contents-oriented networking: i) bundling strategies, ii) bundling-aware sharing systems in BitTorrent, and iii) implications on content-oriented networking.

## 6. REFERENCES

[1] Bittorrent tracker protocol. `http://wiki.theory.org/BitTorrent_Tracker_Protocol`.

[2] The impact of p2p file sharing, voice over ip, instant messaging, one-click hosting and media streaming on the internet. `http://www.ipoque.com/resources/internet-studies/inte`

[3] Open sourced bittorrent client, vuze. `http://www.vuze.com`.

[4] Peer exchange. `http://en.wikipedia.org/wiki/Peer_exchange`.

[5] The pirate bay. `http://thepiratebay.org/`.

[6] A. Carzaniga, M. J. Rutherford, and A. L. Wolf. A routing scheme for content-based networking. In *IEEE INFOCOM*, 2004.

[7] B. Cohen. Incentives build robustness in bittorrent. In *1st Workshop on Economics of Peer-to-Peer Systems*, 2003.

[8] G. Dan and N. Carlsson. Power-law revisited: A large scale measurement study of p2p content popularity. In *USENIX IPTPS*, 2010.

[9] M. Gritter and D. R. Cheriton. An architecture for content routing support in the internet. In *the 3rd Usenix Symposium on Internet Technologies and Systems (USITS)*, 2001.

[10] L. Guo, S. Chen, Z. Xiao, E. Tan, X. Ding, and X. Zhang. Measurements, analysis, and modeling of bittorrent-like systems. In *ACM IMC*, 2005.

[11] J. Han, T. Chung, H. Kim, T. T. Kwon, and Y. Choi. Systematic suppor for content bundling

in bittorrent swarming. In *IEEE INFOCOM Student Workshop*, 2010.

[12] V. Jacobson, D. K. Smetters, J. D. Thornton, M. F. Plass, N. H. Briggs, and R. L. Braynard. Networking named content. In *ACM CoNEXT*, 2009.

[13] M. Kabir. Similarity matching techniques for fault diagnosis in automotive infotainment electronics. *CoRR*, abs/0909.2375, 2009.

[14] T. Koponen, M. Chawla, B.-G. Chun, A. Ermolinskiy, K. H. Kim, S. Shenker, and I. Stoica. A data-oriented (and beyond) network architecture. In *ACM SIGCOMM*, 2007.

[15] A. Legout, N. Liogkas, E. Kohler, and L. Zhang. Clustering and sharing incentives in bittorrent systems. In *ACM SIGMETRICS*, 2007.

[16] D. S. Menasche, G. Neglia, D. Towsley, and S. Zilberstein. Strategic reasoning about bundling in swarming systems. In *IEEE GameNets*, 2009.

[17] D. S. Menasche, A. A. Rocha, B. Li, D. Towsley, and A. Venkataramani. Content availability and bundling in swarming systems. In *ACM CoNEXT*, 2009.

[18] M. Piatek, T. Isdal, T. Anderson, A. Krishnamurthy, and A. Venkataramani. Do incentives build robustness in bittorrent. In *USENIX NSDI*, 2007.

[19] D. Qiu and R. Srikant. Modeling and performance analysis of bittorrent-like peer-to-peer networks. In *ACM SIGCOMM*, 2004.

[20] A. Sherman, J. Nieh, and C. Stein. Fairtorrent: bringing fairness to peer-to-peer systems. In *ACM CoNEXT*, 2009.

[21] Y. Tian, D. Wu, and K.-W. Ng. Analyzing multiple file downloading in bittorrent. In *IEEE ICCP*, 2006.

[22] Y. Yang, A. L. H. Chow, and L. Golubchik. Multi-torrent: A performance study. In *IEEE MASCOTS*, 2008.